\documentclass[12pt]{article}
\begin{document}
\setcounter{footnote}{2}





\title{\bf Standard-like Model from $D=4$ Type IIB Orbifolds}

\author{ Huan-Xiong  Yang
\footnote{E-mail:
hxyang@zimp.zju.edu.cn}
\\
{ \sl
Center for Theoretical  Physics, University of Sussex,}\\{\sl
Brighton BN1  9QJ, UK}  \\
{ \sl
Zhejiang Institute of   Modern Physics, Zhejiang University,}\\
{\sl Hangzhou, 310027, China }}

\date{\today}

\maketitle

\begin{abstract}

Based  on  the twisted   R-R  tadpole  cancellation conditions at   the
singularities of $D=4$ Type IIB orbifold $T^6/{\bf Z_3}$, we propose a
new bottom-up approach to  embed standard model with three generations
into string theory.
\end{abstract}

\vskip2pc

\section{Introduction}

The  discovery of D-branes in  string  theory has largely extended our
view on consistent string theory vacua.  D-branes  are the carriers of
Ramond-Ramond  (R-R) charges on which  the open strings are allowed to
end.  Having  open  string sectors  means having   gauge interactions.
Therefore, string  theory with D-branes  in its  spectrum provides new
scenarios for   finding string theory   vacuum  configuration which is
expected to be  much closer to the  observed standard model (SM).   An
interesting class of string theory vacua is the  Type IIB orbifolds in
$D=4$  dimensions.  They provide  explicit  string theory models where
gauge fields and charged matter   are localized on D-branes,   whereas
gravity  and  other closed  string modes  propagate in full spacetime.
Several    attempts         have      been   made         in      this
direction\cite{9804026}\cite{0004214}\cite{0005067}, among  which  the
standard orientifolds belong to   top-down approach while some   ${\bf
Z_N}$ orbifold models provide possibilities for bottom-up approach. In
all   D-brane  models   strong  consistency    conditions come    from
cancellation  of R-R tadpoles\cite{9601038}\cite{9808139}.  It is also
expected in view of low-energy phenomenology  that the twisted tadpole
cancellation  can   guarantee  the   cancellation of nonabelian  gauge
anomalies.

Among the known D-brane constructions of $D=4$ Type IIB orbifolds, the
``bottom-up  approach''  pioneered in Ref.\cite{0005067}  appears very
interesting.     By locating six  D3-branes   at  $one$ singularity of
$T^6/{\bf Z_3}$  orbifold  of Type  IIB string  theory and including a
stack of   D7-branes in the   configuration  as well,  the authors  of
\cite{0005067}  found a nontrivial    solution   of the  R-R   tadpole
cancellation  conditions  which corresponds   to a gauge   group $U(3)
\times U(2) \times U(1)$, very close to that of standard model, living
in  the worldvolumes of D3-branes at  an orbifold singularity. Another
significant characterictic   of   standard model, quark-lepton  family
replication with just $three$ generations, is also ensured by the fact
that  there are   only  three  complex planes   in that   compactified
space. It is undeniable that  such a bottom-up approach is attractive,
however, there are some serious drawbacks. The massless chiral fermion
spectrum of \cite{0005067} contains only  3 fermion states with  three
generations, their hypercharges are also not the  same as those of the
quarks  and leptons  in standard  model.  In addition,  only the gauge
fields  living  in D3-branes were  supposed  to describe the realistic
interactions while there was  no reliable interpretation for the gauge
fields on D7-branes' worldvolumes.

In this   paper, we report  a revised  version of  the above bottom-up
approach,    which may also   provide a   stringy   realization of the
phenomenological approach  proposed in\cite{0004214}.   Enlightened by
the      recent     progress     in    the     intersecting    D-brane
world\cite{9908130}\cite{0011073}, we  now consider such  a  $T^6/{\bf
Z_3}$  orbifold of  Type IIB  string theory (with  D3-branes and three
kinds of oriented D7-branes and anti-D7-branes in its non-perturbative
spectrum) that there is at most $one$ factorial gauge group $U(n) \,(1
\le  n \le 3   )$ embedded on  the  worldvolumes of each kinds  of the
oriented  D-branes. We also  postulate  the presence of an  additional
Wilson line along the $4$-th (complex)  compactified dimension so that
there is no a factorial gauge group $U(3)$ omnipresent on the oriented
D-branes (D7$_{5, (0)}$-branes   in  our notation, See  Section  $2$).
Three-family replication of chiral fermion spectrum is achieved by the
fact that there    are only three  fixed  points  of  the ${\bf  Z_3}$
orbifold   wrapped by D7$_{5,   (0)}$-branes  which can  not  feel the
presence  of the Wilson  line.  Near these  $3$ fixed points the total
gauge group on  the D-brane worldvolumes   is then the  expected $U(3)
\times U(2)  \times  U(1)_{1} \times  U(1)_{0}$.  The  chiral  fermion
states appear  as the open string Ramond  states stretched between two
different  kinds of D-branes.  The requirement  of cancellation of the
twisted R-R tadpoles leads exactly to the massless left-handed fermion
spectrum of standard model  , with correct hypercharge  assignment and
correct family replication (3  generations), and without cubic $SU(3)$
gauge anomaly.  To our surprise, we can further ensure the twisted R-R
tadpole cancellation conditions and the cubic nonabelian gauge anomaly
cancellation  conditions  at the remaining   $24$ ${\bf Z_3}$ orbifold
fixed points without violating   the above-mentioned properties.   The
only  expense is  that we    have  to  include  $12$ extra    massless
left-handed fermion states into our chiral fermion spectrum, which are
located at  $6$ fixed points separately but  form the singlets of both
``colour'' group $SU(3)$ and  ``weak isospin'' group $SU(2)$.  As most
of  the  similar approaches    initiated from the   intersecting brane
configurations, our model is non-supersymmetric.

The  paper is  organized as follows.   In section  $2$, we  define our
D-brane  configuration at the ${\bf  Z_3}$ orbifold origin.  As a kind
of bottom-up approach within Type IIB  string theory, the guidance for
our  defining D3D7  brane  configuration comes  from the corresponding
twisted  R-R tadpole cancellation  conditions\cite{0005067}\cite{ABY1}
and  the gauge group structure   of standard model  of the  low-energy
particle physics.  With the help of ${\bf Z_3}$ orbifold projection on
Chan-Paton wavefunctions and the constraints  from the cancellation of
cubic   $SU(3)$ gauge anomaly,  we obtain  $five$ massless left-handed
Ramond open string fermion states at origin which  turn out to form an
entire family of the  chiral   fermions of  the standard model.     In
section  $3$, we study the  consistency conditions and their solutions
at remaining     orbifold   fixed points.   In    order   to keep  the
phenomenological attraction of  our model, we  suppose that there is a
discrete Wilson   line  along  the   $4$-th   complex   plane  of  the
orbifold. The tadpole cancellation conditions  and their solutions are
given point  by point, which lead to  another two families of standard
model chiral   fermions and $12$   extra massless fermion  states.  In
section $4$, we give a  detailed discussion about $U(1)$ anomalies and
the corresponding cancellation mechanism.  Section  $5$ is left as our
conclusions and some remarks.

\section{Chiral Fermionic Spectrum at Origin}
\renewcommand{\theequation}{2.\arabic{equation}}\setcounter{equation}{0}

We begin   with   the standard $D=4,  {\mathcal   N}=1$ supersymmetric
$T^6/{\mathbf  Z_3}$ orbifold   construction    of Type  IIB    string
theory\cite{0005067}.  A  set of D3-branes is located at
some singularities  (fixed points)  of  the considered  orbifold.  The
configuration  includes some kinds of oriented  D7-branes as well.  The
D7-branes with different orientations are postulated to be independent
of one  another. We  adopt light-cone gauge,  in which  we use complex
planes representing the  target space dimensions: the $second$ complex
plane  corresponds to non-compact dimensions;   the $3$-rd, $4$-th and
$5$-th   complex  planes correspond  to  ${\bf   Z_3}$ invariant $T^6$
lattice.  The  ${\bf Z_3}$  orbifold action on  open  string states is
given by a matrix
\begin{equation} \label{eq: 101}
\theta =   \textrm{diag}[\,\exp({2 \pi i   a_2}/3), \,  \exp({2  \pi i
a_3}/3), \, \exp({2 \pi i a_4}/3), \, \exp({2 \pi i a_5}/3)\,]
\end{equation}
for worldsheet fermions and
\begin{equation} \label{eq: 102}
\theta = \textrm{diag}[\,\exp({2 \pi i b_3}/3), \, \exp({2 \pi i b_4}/3), \, \exp({2 \pi i b_5}/3)\,]
\end{equation}
for ( complex  ) worldsheet bosons.   In addition, the orbifold action
of the  ${\bf  Z_3}$ generators   $\theta$  must be  embedded  on  the
Chan-Paton     indices,  defined   by diagonal unitary  matrices
$\gamma_{\theta, 3}$  and $\gamma_{\theta, 7}$. In Eqs.(\ref{eq: 101})
and  (\ref{eq:   102})  the twist  parameters   obey  the group theory
constraints $\sum^{5}_{i=2} a_{i}= 0 (\textrm{mod}3)$ and
$$
\left. \begin{array}{lll} b_3 =  - a_4 - a_5, &  \, b_4 = -a_5 -a_3, &
\, b_5 = -a_3 - a_4
\end{array}
\right.
$$
The  ${\mathcal N}=1$ supersymmetry requires condition $\sum^{5}_{r=3}
b_r =0$. Although the model we will consider is non-supersymmetric, we
impose this condition for simplicity. In what follows, we take
\begin{equation} \label{eq: 103}
\left.
\begin{array}{lll}
a_2=0, & \, a_3 =a_4 =1, & \, a_5 = -2\\
       & \, b_3 =b_4 =1, & \, b_5 = -2
\end{array}
\right \}
\end{equation}
to        ensure        a         crystallographical          orbifold
action\cite{9804026}\cite{0005067}.

The ${\bf Z_3}$ invariant $T^6$ can  be realized as  a root lattice of
the Lie algebra $[su(3)]^{3}$.  Let the simple roots
$\{ \vec{\alpha}_{2i-2}$, $\vec{\alpha}_{2i-1} \}$ of $su(3)_{i}$ be the
basis   vectors of  the  $i$-th  complex  plane  of the  $T^6$ lattice
($i=3,\,4,\,5$), the orbifold actions are then equivalently defined by
$\theta     \vec{\alpha}_{2i-2}  =   \vec{\alpha}_{2i-1}$  and $\theta
\vec{\alpha}_{2i-1} = - \vec{\alpha}_{2i-2} - \vec{\alpha}_{2i-1}$. We
can further  express the position  coordinates  of ${\bf  Z_3}$  fixed
points as
\begin{equation} \label{eq:1}
{\bf   x}_{\textrm{f}}     =  \frac{m}{3}   (\vec{\alpha}_{4}     +  2
     \vec{\alpha}_{5})     + \frac{n}{3}      (\vec{\alpha}_{6}   +  2
     \vec{\alpha}_{7})    +    \frac{p}{3}   (\vec{\alpha}_{8}  +    2
     \vec{\alpha}_{9})
\end{equation}
where $m, n,  p =0, \pm 1$. In  what  follows these fixed  points will
alternatively be denoted by $(m, n, p)$.

The local   consistency  of the considered  D-brane   configuration is
guaranteed    by the so-called     twisted  R-R tadpole   cancellation
conditions  at  all   orbifold  fixed   points.   Suppose  that    the
D7$_{3}$-branes carry  negative R-R  charge while the   other D-branes
(namely   D3-,  D7$_4$-    and D7$_5$-branes)    carry   positive  R-R
charge.  Then at the fixed  point $(m, n,  p)$ the twisted R-R tadpole
cancellation conditions for $D=4,  {\mathcal N}=1$ supersymmetric Type
IIB orbifold $T^6/{\mathbf Z_3}$ are\cite{0005067}\cite{ABY1},
\begin{equation} \label{eq: 2}
\left \{
\begin{array}{l}
Tr\gamma_{1, 3,  (m, n, p)} + \frac{1}{3}  Tr\gamma_{1, 7_{3}, (m) } -
\frac{1}{3} Tr\gamma_{1, 7_{4}, (n)} + \frac{1}{3} Tr\gamma_{1, 7_{5},
(p)}=0, \\ Tr\gamma_{2, 3,    (m, n, p)} - \frac{1}{3}    Tr\gamma_{2,
7_{3},  (m)  } + \frac{1}{3} Tr\gamma_{2,   7_{4},  (n)} + \frac{1}{3}
Tr\gamma_{2, 7_{5}, (p)}=0
\end{array}
\right.
\end{equation}
where  we   have used the     similar notation of  Ref.\cite{9804026},
$e. g.$,  $\gamma_{1, 3, (m, n,  p)} \equiv \gamma_{\theta,  3, (m, n,
p)}$  and  $\gamma_{2,  3, (m, n,   p)}  \equiv (\gamma_{1, 3,  (m, n,
p)})^2$. D3$_{(m, n, p)}$ are referred to as  the D3-branes located at
fixed  point $(m, n, p)$, and  D7$_{3,  (m)}$, for example, stands for
D7-branes located at the fixed point $(m)$ on the $3$-rd complex plane
whose worldvolumes are transverse to the $3$-rd complex plane but full
of  the $4$-th   and $5$-th planes.   In   this paper,  we consider  a
non-supersymmetric  version   of     above orbifold,  in    which  the
D7$_{3}$-branes are   replaced    by  the   corresponding  anti-branes
$\widetilde{\textrm  D7}_{3}$. Branes  and anti-branes carry  opposite
charges  with   respect  to the  R-R   fields.  Therefore, instead  of
Eqs.(\ref{eq: 2}),   the   tadpole  cancellation conditions  for   the
considered orbifold should be\cite{9908072}:
\begin{equation} \label{eq: 104}
\left \{
\begin{array}{l}
Tr\gamma_{1, 3,  (m, n, p)} - \frac{1}{3}  Tr\gamma_{1, \tilde{7}_{3}, (m) } -
\frac{1}{3} Tr\gamma_{1, 7_{4}, (n)} + \frac{1}{3} Tr\gamma_{1, 7_{5},
(p)}=0, \\ Tr\gamma_{2, 3,    (m, n, p)} + \frac{1}{3}    Tr\gamma_{2,
\tilde{7}_{3},  (m)  } + \frac{1}{3} Tr\gamma_{2,   7_{4},  (n)} + \frac{1}{3}
Tr\gamma_{2, 7_{5}, (p)}=0
\end{array}
\right.
\end{equation}

The origin  $(0, 0, 0)$ of  $T^6$ lattice is   naturally a ${\bf Z_3}$
fixed point, at which the twisted  R-R tadpole cancellation conditions
are written as,
\begin{equation} \label{eq: 3}
\left \{
\begin{array}{l}
Tr\gamma_{1,  3,  (0,0,0)} - \frac{1}{3}  Tr\gamma_{1,  \tilde{7}_{3}, (0) } -
\frac{1}{3} Tr\gamma_{1, 7_{4}, (0)} + \frac{1}{3} Tr\gamma_{1, 7_{5},
(0)}=0, \\ Tr\gamma_{2, 3,  (0,0,0)} + \frac{1}{3} Tr\gamma_{2, \tilde{7}_{3},
(0) }   +  \frac{1}{3}   Tr\gamma_{2,  7_{4},    (0)}  +   \frac{1}{3}
Tr\gamma_{2, 7_{5}, (0)}=0
\end{array}
\right.
\end{equation}
As   a  starting point of   our  bottom-up  approach,  we consider the
following solutions of Eqs.(\ref{eq: 3}),
\begin{equation} \label{eq: 4}
\left.
\begin{array}{llr}
\gamma_{1, 3, (0, 0,  0)} &  =  & {\bf 1}_{1}  \\
\gamma_{1, \tilde{7}_{3}, (0)}    &  =  &  - \alpha {\bf 1}_{1} \\
\gamma_{1, 7_{4}, (0)}    &  =  &  - \alpha {\bf 1}_{2} \\
\gamma_{1, 7_{5}, (0)}    &  =  & \alpha^2 {\bf 1}_{3}
\end{array}
\right.
\end{equation}
where $\alpha = \exp[\,{(2 \pi i)}/{3}\,]$.
This  assignment  leads to the  gauge  group $U(3)  \times U(2) \times
U(1)_{1,(0)} \times U(1)_{0,(0, 0, 0)}$  living in the worldvolumes of
D-branes near origin:
\begin{equation} \label{eq: 5}
\left.
\begin{array}{ll}
\textrm{D3}_{(0, 0, 0)}: &  U(1)_{0,(0, 0, 0)} \\
\widetilde{\textrm{D7}}_{3, (0)}: &  U(1)_{1, (0)} \\
\textrm{D7}_{4, (0)}: &  U(2) = U(1)_{2, (0)} \times SU(2) \\
\textrm{D7}_{5, (0)}: &  U(3) = U(1)_{3} \times SU(3)
\end{array}
\right.
\end{equation}

We now   consider the massless  chiral fermion  states in all possible
open string sectors at origin.

\subsection{\bf{33}, \bf{$\tilde{\textrm 7}$}$_3$\bf{$\tilde{\textrm 7}$}$_3$, \bf{7$_4$7$_4$} and \bf{7$_5$7$_5$} sectors}

In $\bf{33}$  sector,  the massless Ramond  fermions are  of the  form
$\vert  s_2 s_3 s_4 s_5   \rangle$ (in light-cone  gauge), where every
$s_i$  is either of $\pm \frac{1}{2}$.   The value of $s_2$ determines
the spacetime fermion chirality. The  worldsheet fermion number can be
defined as $(-1)^F =   \exp[i \pi(\sum_{i=2}^{5}s_i)]$ so that  we can
implement   $\mathrm{GSO}$  projection    $(-1)^F  =1$   by    letting
$\sum^{5}_{i=2}  s_i=$ even.  Before  $\bf{Z}_3$ orbifold  projection,
the  left-handed  massless  states  ($s_2=-\frac{1}{2}$) are explictly
stated below,
\begin{equation} \label{eq: 6}
\left.
\begin{array} {llrrrrr}
|\Psi_1  \rangle &  = &  |  & -   \frac{1}{2}, &  -  \frac{1}{2}, &  -
\frac{1}{2},  & - \frac{1}{2} \rangle  \\ |\Psi_2 \rangle &  = & | & -
\frac{1}{2}, & - \frac{1}{2}, & \frac{1}{2}, & \frac{1}{2} \rangle \\
|\Psi_3   \rangle &  =  & |   &  -  \frac{1}{2}, &  \frac{1}{2},   & -
\frac{1}{2}, &  \frac{1}{2} \rangle  \\ |\Psi_4 \rangle   & = &  | & -
\frac{1}{2}, & \frac{1}{2}, & \frac{1}{2}, & - \frac{1}{2} \rangle
\end{array}
\right.
\end{equation}
Under orbifold projection,
$$
\left.
\begin{array}{ll}
\theta^k |s_{2}s_{3}s_{4}s_{5}\rangle =   \exp[\frac{2  \pi   i  k}{3}
(   s_{2}a_{2}   +     s_{3}a_{3}  +   s_{4}a_{4}    +    s_{5}a_{5})]
|s_{2}s_{3}s_{4}s_{5}\rangle \, \, \, \, \, & (k=1, 2)
\end{array}
\right.
$$
Explicitly,
\begin{eqnarray} \label{eq: 7}
& \theta |\Psi_1  \rangle & = |  \Psi_1 \rangle \nonumber  \\ & \theta
|\Psi_j \rangle & = \alpha | \Psi_j \rangle \, \, \, (j=2, 3, 4)
\end{eqnarray}
The open string spectrum is constructed by requiring the states $\vert
\Psi, ij \rangle \lambda_{ji}$ to be invariant under the action of the
orbifold projection. As a result, the  Chan-Paton wavefunctions of the
above  massless   open  string states are    determined  by conditions
\cite{0005067},
\begin{equation} \label{eq: 8}
\left.
\begin{array}{lll}
\lambda_1  = \gamma_{1, 3, (0,  0, 0)} \lambda_1  \gamma_{1, 3, (0, 0,
0)}^{-1} , \,   \, \,& \lambda_j =  \alpha  \gamma_{1, 3, (0,   0, 0)}
\lambda_j \gamma_{1, 3, (0, 0, 0)}^{-1} & (j=2, 3, 4)
\end{array}
\right.
\end{equation}
This  leads to $one$   massless left-handed fermion state which  forms
the adjoint representation of Abelian gauge group $U(1)_{0, (0, 0, 0)}$.

Similar situations appear in other three  open string sectors ${\tilde
{\bf 7}}_3 {\tilde {\bf 7}}_3$, ${\bf 7}_4{\bf7}_4$ and ${\bf 7}_5{\bf
7}_5$.  Each of these sectors  has such a left-handed massless fermion
state  that forms  the  adjoint  representation of the   corresponding
unitary group.   Actually, this   is  also true  for the  right-handed
massless fermion states  in  these sectors.  These states  do actually
form the non-chiral   massless   fermions  with all   $U(1)$   charges
vanishing. In supersymmetric models, these states could be interpreted
as the  super-partners  of the $U(1)$ gauge  bosons.    Because of the
non-chirality  of these states,  We will  simply  ignore them from the
discussion about the chiral fermion spectrum.
%
%

\subsection{3$\tilde{\textrm 7}_{3}$ and $\tilde{\textrm 7}_{3}$3 sectors}

Before orbifold  projection, the massless  Ramond states  in these two
sectors  are $\vert   s_2 s_3   \rangle$.   Relying on  the fact  that
$\widetilde{\textrm D7}_3$ stands  for  an anti-brane, we  should take
$(-1)^F =-1$  as   the $\mathrm{GSO}$ projection\cite{9908072}.   This
implies that $s_2 = s_3$.   There is only $one$ left-handed  candidate
($s_2= - \frac{1}{2}$) in each sector,
\begin{equation} \label{eq: 9}
\vert \Psi \rangle = \vert -\frac{1}{2}, -\frac{1}{2} \rangle
\end{equation}
which undergoes the following ${\bf Z_3}$ orbifold action,
\begin{equation} \label{eq: 10}
\theta \vert \Psi \rangle = \exp(-\frac{\pi  i}{3}) \vert \Psi \rangle
= - \alpha \vert \Psi \rangle
\end{equation}
Therefore,
\begin{equation} \label{eq: 11}
\lambda = -  \alpha \gamma_{1, 3, (0,  0,  0)} \lambda \gamma^{-1}_{1,
7_3,  (0)} = - \alpha \gamma_{1,  7_3, (0)} \lambda \gamma^{-1}_{1, 3,
(0, 0, 0)}
\end{equation}
Eqs.(\ref{eq:  11})  lead  to a massless   Ramond   fermionic state in
${\bf3}{\bf   {\tilde  7}_3}$  sector,   which  forms  the fundamental
representation   of   $U(1)_{0, (0,0,   0)}$ and  the anti-fundamental
representation of $U(1)_{1, (0)}$.
%
%

\subsection{$\tilde{\textrm 7}_3$7$_4$ and 7$_4$$\tilde{\textrm 7}_3$ sectors}

The massless Ramond fermionic states before orbifold action are $\vert
s_2  s_5   \rangle$.  The $\mathrm{GSO}$   projection  $(-1)^{F}= - 1$ is
implemented by  setting $s_2 = s_5$  and we  have only one left-handed
candidate in each sector,
\begin{equation} \label{eq: 12}
\vert \Psi \rangle = \vert -\frac{1}{2}, -\frac{1}{2} \rangle
\end{equation}
Under the ${\bf Z_3}$ orbifold projection,
\begin{equation} \label{eq: 13}
\theta \vert \Psi \rangle = \exp(\frac{2 \pi i}{3}) \vert \Psi \rangle
= \alpha \vert \Psi \rangle
\end{equation}
Then,
\begin{equation} \label{eq: 14}
\lambda =  \alpha \gamma_{1,  7_3,  (0)} \lambda  \gamma^{-1}_{1, 7_4,
(0)} = \alpha \gamma_{1, 7_4, (0)} \lambda \gamma^{-1}_{1, 7_3, (0)}
\end{equation}
There  is no qualified massless left-handed  Ramond state in either of
these two sectors obeying Eqs.(\ref{eq: 14}).

\subsection{${\tilde{\textrm 7}}_3$7$_5$ and 7$_{5}$${\tilde{\textrm 7}}_3$ sectors}

The  massless Ramond  fermionic   states before ${\bf Z_3}$   orbifold
action  are $\vert s_2  s_4   \rangle$. The $\mathrm{GSO}$  projection
$(-1)^{F}= - 1$ is implemented  by setting $s_2 = s_4$  and we  have only
one left-handed candidate in each sector,
\begin{equation} \label{eq: 15}
\vert \Psi \rangle = \vert -\frac{1}{2}, -\frac{1}{2} \rangle
\end{equation}
Under the ${\bf Z_3}$ orbifold projection,
\begin{equation} \label{eq: 16}
\theta \vert \Psi \rangle  = \exp(-\frac{\pi i}{3}) \vert \Psi \rangle
= - \alpha \vert \Psi \rangle
\end{equation}
Then,
\begin{equation} \label{eq: 17}
\lambda = - \alpha \gamma_{1,  7_3, (0)} \lambda \gamma^{-1}_{1,  7_5,
(0)} = - \alpha \gamma_{1, 7_5, (0)} \lambda \gamma^{-1}_{1, 7_3, (0)}
\end{equation}
Eqs.(\ref{eq: 17}) lead to a  massless Ramond fermionic state in
${\tilde {\bf 7}}_3 {\bf 7}_5$ sector. It forms the fundamental representation of $U(1)_{1,
(0)}$ and the anti-fundamental representation of nonabelian gauge
group $U(3)$.
%
%

\subsection{37$_5$ and 7$_5$3 sectors}

Before orbifold projection, the  massless  Ramond fermions are  $\vert
s_2 s_5 \rangle$. These fermions are the Ramond open strings stretched
between  two D-branes rather than brane-antibrane  pairs. So we should
take $(-1)^{F}=  1$ as $\mathrm{GSO}$  projection in these sectors. In
fact, such a $\mathrm{GSO}$   projection  is also the requirement   of
cancellation of  cubic $SU(3)$  gauge  anomaly at  origin.  Obviously,
$(-1)^{F}=  1$ in the  considered sectors can  be achieved via setting
$s_2 = - s_5$.  As before, there is only one left-handed
$(s_2 =- \frac{1}{2})$ candidate in each sector,
\begin{equation} \label{eq: 18}
\vert \Psi \rangle = \vert -\frac{1}{2}, \frac{1}{2} \rangle
\end{equation}
on which the ${ \bf Z_3}$ orbifold projection acts as,
\begin{equation} \label{eq: 19}
\theta  \vert   \Psi \rangle =  \exp(-\frac{2   \pi i}{3})  \vert \Psi
\rangle = \alpha^2 \vert \Psi \rangle
\end{equation}
So,
\begin{equation} \label{eq: 20}
\lambda = \alpha^2 \gamma_{1,  3,  (0, 0, 0)} \lambda  \gamma^{-1}_{1,
7_5, (0)}  = \alpha^2 \gamma_{1, 7_5,  (0)} \lambda \gamma^{-1}_{1, 3,
(0, 0, 0)}
\end{equation}
Eqs.(\ref{eq: 20}) lead to a massless  Ramond fermion state in ${\bf
37_5}$ sector. The state forms the fundamental representation of
$U(1)_{0, (0,0,0)}$ and the anti-fundamental representation of $U(3)$.
%
%

\subsection{7$_4$7$_5$ and 7$_5$7$_4$ sectors}

Before orbifold  projection,   the massless ramond  fermions  in these
sectors are $\vert s_2 s_3 \rangle$.   as explained in last paragraph,
we  have to  set $s_2  = -  s_3$  realizing  $\mathrm{GSO}$ projection
$(-1)^{F}= 1$.   There is only  one left-handed $(s_2 =- \frac{1}{2})$
candidate in each sector,
\begin{equation} \label{eq: 21}
\vert \Psi \rangle = \vert -\frac{1}{2}, \frac{1}{2} \rangle
\end{equation}
on which the ${ \bf Z_3}$ orbifold projection acts as,
\begin{equation} \label{eq: 22}
\theta \vert \Psi \rangle =  \exp(\frac{ \pi i}{3}) \vert \Psi \rangle
= - \alpha^2 \vert \Psi \rangle
\end{equation}
Therefore,
\begin{equation} \label{eq: 23}
\lambda = - \alpha^2 \gamma_{1, 7_4, (0)} \lambda \gamma^{-1}_{1, 7_5,
(0)} =  - \alpha^2 \gamma_{1,  7_5, (0)}  \lambda \gamma^{-1}_{1, 7_4,
(0)}
\end{equation}
Eqs.(\ref{eq: 23}) lead to a  massless Ramond fermionic state in ${\bf
7_57_4}$ sector which forms the fundamental representation of gauge
group $U(3)$ but the anti-fundamental representation of $U(2)$.
%
%

\subsection{37$_4$ and 7$_4$3 sectors}

In these two sectors,  the  massless ramond  fermions  before $ {  \bf
Z_3}$ orbifold projection are  of the  form  $\vert s_2 s_4  \rangle$.
The $\mathrm{GSO}$ projection $(-1)^{F}= 1$ is implemented by setting $s_2 = - s_4$.
There is  only one left-handed $(s_2  =- \frac{1}{2})$  candidate in
each sector,
\begin{equation} \label{eq: 24}
\vert \Psi \rangle = \vert -\frac{1}{2}, \frac{1}{2} \rangle
\end{equation}
on which the ${\bf Z}_3$ orbifold projection acts as,
\begin{equation} \label{eq: 25}
\theta \vert \Psi \rangle = \exp(\frac{  \pi i}{3}) \vert \Psi \rangle
= - \alpha^2 \vert \Psi \rangle
\end{equation}
Therefore,
\begin{equation} \label{eq: 26}
\lambda = - \alpha^2 \gamma_{1, 3,  (0, 0, 0)} \lambda \gamma^{-1}_{1,
7_4, (0)} = - \alpha^2 \gamma_{1, 7_4, (0)} \lambda \gamma^{-1}_{1, 3,
(0, 0, 0)}
\end{equation}
They give rise to a  massless Ramond fermion  state in ${\bf 7_4 3}$
sector, forming the fundamental representation of gauge group $U(2)$
while the anti-fundamental representation $U(1)_{0, (0, 0, 0)}$.
%
%

Consequently,   we obtain   exactly one-family  Standard-Model  chiral
spectrum from open   strings stretching between  different  D-branes at
${\bf Z_3}$ orbifold origin:
\begin{center}
\begin{tabular}{|c|c|c|c|r|r|r|r|}\hline
\multicolumn{8}{|c|}{\bfseries{Chiral Fermionic States at ${\bf Z}_3$
Orbifold Origin}}\\
\hline State & Sector & Rep. of SU(3) & Rep. of
SU(2) & Q$_{3}\,\,\,$ & Q$_{2, (0)}\,\,\,$ & Q$_{1, (0)}$ & Q$_{0, {(0, 0, 0)}}$ \\
\hline (${\bf 3, \bar2}$) & 7$_5$7$_4$ & ${\bf 3}$ & ${\bf {\bar 2}}$ & 1 & -1 & 0 &
0 \\
\hline (${\bf \bar3, 1}$)$^{'}$ & ${\tilde 7}_3$7$_5$ & ${\bf {\bar 3}}$ & ${\bf 1}$ & - 1 & 0 & 1 &
0 \\
\hline(${\bf \bar3, 1}$) & 37$_5$ & ${\bf {\bar 3}}$ & ${\bf 1}$ & - 1 & 0 & 0 &
1 \\
\hline (${\bf 1, 2}$) & 7$_4$3 & ${\bf 1}$ & ${\bf 2}$ & 0 & 1 & 0 &
-1 \\
\hline(${\bf 1, 1}$) & 3${\tilde 7}_3$ & ${\bf 1}$ & ${\bf 1}$ & 0 & 0 & -1 &
1 \\
\hline
\end{tabular}
\end{center}

\section{Chiral Fermionic Spectrum at Other ${\bf Z_3}$ singularities}
\renewcommand{\theequation}{3.\arabic{equation}}\setcounter{equation}{0}

\subsection{Tadpole cancellation conditions}

In the last section, we only discussed the cancellation of the twisted
R-R tadpoles  at ${ \bf Z_3}$ orbifold  origin.  Because each oriented
D7-brane  wraps  nine fixed  points, we   have to  ensure  the tadpole
cancellation at all  of them. The simplest  way to such a cancellation
is  by a  replication at every  orbifold  fixed point  of the  D-brane
configuration at the origin.    This, however, obviously  violates the
phenomenological attraction  of  the considered  orbifold because  the
standard   model  only     contains  $3$  quark-lepton    generations.
Fortunately,   in the model   under  consideration, three  families of
massless chiral fermion states can be obtained by introducing a Wilson
line $e^{i \oint A \cdot dl}$ passing the orbifold origin and along
the   $4$-th complex plane.    The   presence of such  a  Wilson  line
introduces  probably    additional   gauge   field   structure  on the
worldvolumes of  the ${\widetilde{\textrm  D7}}_{3, (0)}$ and
${\textrm D7}_{5, (0)}$ branes, which will affect the parallel transport
along the $4$-th complex plane. The $6$ fixed points $ (0, \pm 1, p) $
with  $p=0, \pm1$ wrapped by  ${\widetilde {\textrm D7}}_{3, (0)}$ and
$6$ fixed points $ (m, \pm 1, 0) $ with $m=0, \pm 1$ wapped by D7$_{5,
(0)}$ can feel the presence  of this  Wilson line.  Nevertheless,  the
fixed  points ($m, 0, p$) wrapped  by D7$_{4,  (0)}$  can not feel its
presence ($m$ and $p$ do not vanish at the same  time), either can not
the   fixed  points ($0, 0, \pm    1$) wrapped by ${\widetilde{\textrm
D7}}_{3,  (0)}$-branes   and  ($\pm  1, 0,    0$)  wrapped by  D7$_{5,
(0)}$-branes.    Therefore,  the   twisted  R-R   tadpole cancellation
conditions  at the remaining  ${\bf Z}_3$ fixed points are categorized
into several classes. We list them point by point:

{\bf At Fixed Points ($\pm1, 0, 0$)}:
\begin{equation} \label{eq: 27}
 Tr\gamma_{1, 3, (\pm1, 0, 0)}  - \frac{1}{3} Tr\gamma_{1, {\tilde 7}_3, (\pm1)}
-\frac{1}{3}  Tr\gamma_{1, 7_4,  (0)} + \frac{1}{3}  Tr\gamma_{1, 7_5,
(0)}  =   0
\end{equation}

{\bf At Fixed Points ($0, \pm1, 0$)}:
\begin{equation} \label{eq: 28}
Tr\gamma_{1, 3, (0, \pm1, 0)}
-  \frac{1}{3} Tr[\gamma_{1, {\tilde 7}_3, (0)}\gamma^{\pm 1}_{W4,
{\tilde 7}_3}]
-  \frac{1}{3} Tr\gamma_{1, 7_4, (\pm 1)}
+ \frac{1}{3} Tr[\gamma_{1, 7_5, (0)}\gamma^{\pm 1}_{W4, 7_5}]   =  0
\end{equation}

{\bf At Fixed Points ($0, 1, \pm1 $) and ($0, -1, \pm 1$)}:
\begin{equation} \label{eq: 29}
\left \{
\begin{array}{l}
Tr\gamma_{1, 3, (0, \,\,1, \pm 1)}
- \,\, \frac{1}{3} Tr[\gamma_{1, {\tilde 7}_3, (\pm1)}\gamma_{W4,
{\tilde 7}_3}]
- \,\, \frac{1}{3} Tr\gamma_{1, 7_4, (\,\,1)}
+ \frac{1}{3} Tr\gamma_{1, 7_5, (\pm 1)}  =  0 \\
Tr\gamma_{1, 3, (0, -1, \pm 1 )}
-  \frac{1}{3} Tr[\gamma_{1, {\tilde 7}_3, (\pm 1)}\gamma^{- 1}_{W4,
{\tilde 7}_3}]
-  \frac{1}{3} Tr\gamma_{1, 7_4, (- 1)}
+ \frac{1}{3} Tr\gamma_{1, 7_5, (\pm 1)}   =  0
\end{array}
\right.
\end{equation}

{\bf At Fixed Points ($m, 0, \pm1$) with $m=0, \pm1$}:
\begin{equation} \label{eq: 31}
Tr\gamma_{1, 3, (m, 0, \pm1)}
-  \frac{1}{3} Tr\gamma_{1, {\tilde 7}_3, (m)}
-  \frac{1}{3} Tr\gamma_{1, 7_4, (0)}
+ \frac{1}{3} Tr\gamma_{1, 7_5, (\pm1)}   =  0, \,\,\,
\end{equation}

{\bf At Fixed Points ($\pm1, 1, 0$) and ($\pm 1, -1, 0$)}:
\begin{equation} \label{eq: 32}
\left \{
\begin{array}{l}
Tr\gamma_{1, 3, (\pm1,\,\, 1, 0)}
- \,\, \frac{1}{3} Tr\gamma_{1, {\tilde 7}_3, (\pm1)}
- \, \frac{1}{3} Tr\gamma_{1, 7_4, ( 1)}
 \, + \,\, \frac{1}{3} Tr[\gamma_{1, 7_5, (0)}\gamma_{W4, 7_5}]  =  0
\\
Tr\gamma_{1, 3, (\pm1, -1, 0)}
-  \frac{1}{3} Tr\gamma_{1, {\tilde 7}_3, (\pm1)}
-  \frac{1}{3} Tr\gamma_{1, 7_4, (- 1)}
+ \frac{1}{3} Tr[\gamma_{1, 7_5, (0)}\gamma^{- 1}_{W4, 7_5}]  =  0
\end{array}
\right.
\end{equation}

{\bf At Fixed Points ($m, n, p$) with $ m, n, p$ = $\pm 1$}:
\begin{equation} \label{eq: 34}
\left.
\begin{array}{ll}
Tr\gamma_{1, 3, (m, n, p)}
-  \frac{1}{3} Tr\gamma_{1, {\tilde 7}_3, (m)}
-  \frac{1}{3} Tr\gamma_{1, 7_4, (n)}
+ \frac{1}{3} Tr\gamma_{1, 7_5, (p)}   =  0,
&
\end{array}
\right.
\end{equation}

Let the Wilson line action on Chan-Paton wavefunctions be given by the
following diagonal unitary matrices\cite{0010091, 0006049}:
\begin{equation}\label{eq: 35}
\left \{
\begin{array}{l}
\gamma_{W4, {\tilde 7}_3}= {\bf 1}_{1} \\ \gamma_{W4, 7_5}=\textrm{diag}
({\bf 1}_1, \alpha {\bf 1}_1, \alpha^2 {\bf 1}_1 )
\end{array}
\right.
\end{equation}
This   assignment defines  nontrivial gauge   field  structure on  the
worldvolumes of  D7$_{5, (0)}$-branes: at the  six  fixed points $ (m,
\pm 1,  0) $ with $m=0,   \pm 1$ which can  feel  the presence  of the
Wilson  line  the full gauge  group   is reduced to $U(1)_{31}  \times
U(1)_{32} \times U(1)_{33}$. Only at the  remaining three fixed points
$(m, 0, 0)$ with $m=0, \pm 1$, which can not  feel the presence of the
Wilson line, does the gauge group on the D7$_{5, (0)}$-branes turn out
to become the  ``colour'' group $U(3)$.  Consequently, there is  not a
gauge  group   $U(3)$   omnipresent on  the   worldvolumes  of D7$_{5,
(0)}$-branes.

With the above  Wilson  line, we can  achieve the  cancellation of the
twisted R-R tadpoles at all of the ${\bf  Z_3}$ orbifold fixed points.
The corresponding solutions  for the non-vanishing  gamma matrices are
as follows:
\begin{equation} \label{eq: 36}
\left.
\begin{array}{llrl}
\gamma_{1, 3, (\pm 1, 0, 0)} &  = &  {\bf 1}_1 &\\
\gamma_{1, 3, (m, 0, \pm 1)} &  = &  - \alpha {\bf 1}_1  &\,\,\,\,\, (m=0, \, \pm 1) \\
\gamma_{1, {\tilde 7}_3, (\pm 1) }    &  = &  - \alpha {\bf 1}_1  &\,\\
\gamma_{1, 7_4, (\pm 1) }    &  = &  \alpha {\bf 1}_1   &
\end{array}
\right.
\end{equation}
The other gamma matrices are supposed to be vanishing.

\subsection{Gauge group and chiral fermion spectrum at fixed points
($\pm 1, 0, 0$)}

The twisted R-R tadpole cancellation conditions have been given in
Eqs.(\ref{eq: 27}). According the assignments in Eqs.(\ref{eq: 4}) and
(\ref{eq: 36}), we have the solutions:
\begin{equation} \label{eq: 37}
\left.
\begin{array}{llr}
\gamma_{1, 3, (\pm 1, 0,  0)} &  = &  {\bf 1}_{1}  \\
\gamma_{1, {\tilde 7}_{3}, (\pm 1)}    &  = &  - \alpha {\bf 1}_{1} \\
\gamma_{1, 7_{4}, (0)}        &  = &  - \alpha {\bf 1}_{2} \\
\gamma_{1, 7_{5}, (0)}        &  = &    \alpha^2 {\bf 1}_{3}
\end{array}
\right.
\end{equation}
which imply that the gauge groups
$$
U(3) \times U(2) \times U(1)_{1, (\pm 1)} \times
U(1)_{0, (\pm 1, 0, 0)}
$$
have  been   arranged   in  the   worldvolumes of   the  D3-, D7-  and
${\widetilde{\textrm  D7}}$-branes   near   these  two  fixed  points.
$U(1)_{0, (\pm 1, 0, 0)}$ correspond to  the gauge fields on D3-branes
at fixed points $(\pm1,  0, 0)$ and $U(1)_{1, (\pm  1)}$ to the  gauge
fields  on ${\widetilde{\textrm  D7}}_{3,  (\pm 1)}$-branes.  They are
independent of   one another and  have nothing  to do  with  the gauge
groups near origin.

The similarities between  solutions  (\ref{eq: 37}) and  (\ref{eq: 4})
result   in the conclusion that  there  are two families of Standard
Model left-handed massless fermionic states located at these two fixed
points, and  one  family at  each point.  We summarize  the massless
chiral fermionic spectra into the following two tables:

{\bf At Fixed point ($1, 0, 0$)}:
\begin{center}
\begin{tabular}{|c|c|c|c|r|r|r|r|}\hline
\multicolumn{8}{|c|}{\bfseries{Chiral Fermionic States at ${\bf Z}_3$
Orbifold Fixed Point (1, 0, 0)}}\\
\hline State & Sector & Rep. of SU(3) & Rep. of
SU(2) & Q$_{3}\,\,\,$ & Q$_{2, (0)}\,\,\,$ & Q$_{1, (1)}$ & Q$_{0, (1,
0, 0)}$ \\
\hline (${\bf 3, \bar2}$) & 7$_5$7$_4$ & ${\bf 3}$ & ${\bf {\bar 2}}$ & 1 & -1 & 0 &
0 \\
\hline (${\bf \bar3, 1}$)$^{'}$ & ${\tilde 7}_3$7$_5$ & ${\bf {\bar 3}}$ & ${\bf 1}$ & - 1 & 0 & 1 &
0 \\
\hline(${\bf \bar3, 1}$) & 37$_5$ & ${\bf {\bar 3}}$ & ${\bf 1}$ & - 1 & 0 & 0 &
1 \\
\hline (${\bf 1, 2}$) & 7$_4$3 & ${\bf 1}$ & ${\bf 2}$ & 0 & 1 & 0 &
-1 \\
\hline(${\bf 1, 1}$) & 3${\tilde 7}_3$ & ${\bf 1}$ & ${\bf 1}$ & 0 & 0 & -1 &
1 \\
\hline
\end{tabular}
\end{center}

{\bf At Fixed point ($-1, 0, 0$)}:
\begin{center}
\begin{tabular}{|c|c|c|c|r|r|r|r|}\hline
\multicolumn{8}{|c|}{\bfseries{Chiral Fermionic States at ${\bf Z}_3$
Orbifold Fixed Point (-1, 0, 0)}}\\
\hline State & Sector & Rep. of SU(3) & Rep. of
SU(2) & Q$_{3}\,\,\,$ & Q$_{2, (0)}\,\,\,$ & Q$_{1, (-1)}$ & Q$_{0,
(-1, 0, 0)}$ \\
\hline (${\bf 3, \bar2}$) & 7$_5$7$_4$ & ${\bf 3}$ & ${\bf {\bar 2}}$ & 1 & -1 & 0 &
0 \\
\hline (${\bf \bar3, 1}$)$^{'}$ & ${\tilde 7}_3$7$_5$ & ${\bf {\bar 3}}$ & ${\bf 1}$ & - 1 & 0 & 1 &
0 \\
\hline(${\bf \bar3, 1}$) & 37$_5$ & ${\bf {\bar 3}}$ & ${\bf 1}$ & - 1 & 0 & 0 &
1 \\
\hline (${\bf 1, 2}$) & 7$_4$3 & ${\bf 1}$ & ${\bf 2}$ & 0 & 1 & 0 &
-1 \\
\hline(${\bf 1, 1}$) & 3${\tilde 7}_3$ & ${\bf 1}$ & ${\bf 1}$ & 0 & 0 & -1 &
1 \\
\hline
\end{tabular}
\end{center}

So   far,  we have exactly  obtained   the correct left-handed fermion
spectrum of Standard Model  with  {\bf{three}} generations.  The  weak
hypercharge is defined as
\begin{equation} \label{eq: 38}
Y =  \frac{2}{3} Q_3 +  \frac{1}{2}  Q_{2,(0)} + \sum_{m=0,  \, \pm 1}
Q_{0, (m, 0, 0)}
\end{equation}
The following is the hypercharge distribution in one family of the
massless chiral fermion spectrum:
\begin{center}
\begin{tabular}{|c|c|c|c|c|c|} \hline
\multicolumn{6}{|c|}{\bfseries{Hypercharges in One Family }} \\ \hline
 State  & (${\bf 3, \bar{2}}$) & (${\bf \bar{3}, 1}$)$^{'}$ &
 (${\bf \bar{3}, 1}$) & (${\bf 1, 2}$) & (${\bf 1, 1}$) \\
\hline $Y$  &  1/6 & - 2/3 & 1/3 & - 1/2 & 1 \\ \hline
\end{tabular}
\end{center}

\subsection{Gauge group and chiral fermion spectrum at fixed points
($0, \pm 1, 0$)}

The  twisted R-R tadpole cancellation conditions   have been listed in
Eqs.(\ref{eq: 28}). Corresponding to the assignment in (\ref{eq: 36}),
we have the following gamma matrices at these fixed points:
\begin{equation} \label{eq: 39}
\left.
\begin{array}{lll}
\gamma_{1, 3, (0, \pm 1,  0)} &  = & {\bf 0}  \\
\gamma_{1, {\tilde 7}_{3}, (0)}\gamma^{\pm 1}_{W4, {\tilde 7}_3} & = & - \alpha {\bf 1}_{1} \\
\gamma_{1, 7_{4}, (\pm 1)} &  = &  \alpha {\bf 1}_{1} \\
\gamma_{1, 7_{5}, (0)}\gamma_{W4, 7_5} &  =  & \textrm{diag}
(\alpha^2 {\bf 1}_{1}, {\bf 1}_{1}, \alpha {\bf 1}_{1} ) \\
\gamma_{1, 7_{5}, (0)}\gamma^{- 1}_{W4, 7_5} &  =  & \textrm{diag}
(\alpha^2 {\bf 1}_{1}, \alpha {\bf 1}_{1},  {\bf 1}_{1} )
\end{array}
\right.
\end{equation}
which correspond to the following assignments of gauge groups
\begin{equation} \label{eq: 40}
\left.
\begin{array}{ll}
\widetilde{\textrm{D7}}^{\textrm{Wilson\, line}}_{3, (0)}:  & U(1)_{1,
(0)} \\ \textrm{D7$_{4,   (\pm    1)}$}:  & U(1)_{2,  (\pm     1)}  \\
\textrm{D7$^{\textrm{Wilson\, line}}_{5, (0)}$}: & U(1)_{31}    \times
U(1)_{32} \times U(1)_{33}
\end{array}
\right.
\end{equation}
on the D-branes' worldvolumes near these  fixed points. The conditions
of determining massless left-handed  fermion  spectrum at these  fixed
points are stated as,
\begin{equation} \label{eq: 41}
\left.
\begin{array}{ll}
\widetilde{\textrm{D7}}_{3}\textrm{D7}_{4}: \,\,\,\, & \lambda \, = \,
\alpha  \, [\,  \gamma_{1,  {\tilde   7}_3, (0)} \gamma_{W4,   {\tilde
7}_3}^{\pm   1} \,] \,  \lambda \,  \gamma^{-1}_{1,  7_4,  (\pm 1)} \\
\textrm{D7}_{4}\widetilde{\textrm{D7}}_{3}: \,\,\,\, & \lambda \, = \,
\alpha  \, \gamma_{1,   7_4,  (\pm 1)} \,  \lambda   \, [\, \gamma_{1,
{\tilde 7}_3, (0)}   \gamma_{W4,  {\tilde 7}_3}^{\pm 1}  \,]^{-1}   \\
\widetilde{\textrm{D7}}_{3}\textrm{D7}_{5}: \,\,\,\, & \lambda\, =\, -
\alpha  \, [\, \gamma_{1,   {\tilde  7}_3,  (0)} \gamma_{W4,   {\tilde
7}_3}^{\pm 1} \,] \, \lambda \, [  \, \gamma_{1, 7_5, (0)} \gamma_{W4,
7_5}^{\pm  1} \, ]^{-1} \\ \textrm{D7}_{5}\widetilde{\textrm{D7}}_{3}:
\,\,\,\, & \lambda\,   =  \, -  \alpha  \,  [\, \gamma_{1,  7_5,  (0)}
\gamma_{W4,  7_5}^{\pm 1}  \, ]  \, \lambda \,  [ \,\gamma_{1, {\tilde
7}_3,  (0)}  \gamma_{W4,     {\tilde   7}_3}^{\pm 1}  \,]^{-1}      \\
\textrm{D7}_{4}\textrm{D7}_{5}:  \,\,\,\, & \lambda\,  = \, - \alpha^2
\, \gamma_{1, 7_4, (\pm  1)} \,  \lambda \,  [\, \gamma_{1, 7_5,  (0)}
\gamma_{W4,  7_5}^{\pm 1} \, ]^{-1} \\ \textrm{D7}_{5}\textrm{D7}_{4}:
\,\,\,\, & \lambda \,   = \, -   \alpha^2 \, [\, \gamma_{1,  7_5, (0)}
\gamma_{W4, 7_5}^{\pm 1} \,]  \, \lambda \, \gamma^{-1}_{1, 7_4,  (\pm
1)}
\end{array}
\right.
\end{equation}
They lead  to two massless  left-handed Ramond fermion states  at each
fixed point. The chiral fermion spectrum can be summarized as,
\begin{center}
\begin{tabular}{|r|r|r|r|r|r|r|r|} \hline
\multicolumn{8}{|c|}{\bfseries{Massless Chiral Fermionic Spectra at
($0, \pm 1, 0$) }} \\ \hline
 Fixed Point & Sector  & State & Q$_{31}$ & Q$_{32}$ & Q$_{33}$ & Q$_{2, (\pm 1)}$ & Q$_{1, (0)}$ \\
\hline
 (0, 1, 0) & ${\tilde 7}_3$7$_5$  & ({\bf 1, 1}) & - 1 & 0 & 0 & 0 & 1 \\
\hline
 (0, 1, 0) & 7$_5$${\tilde 7}_3$  & ({\bf 1, 1}) &   0 & 1 & 0 & 0 & -1 \\
\hline
 (0, -1, 0) & ${\tilde 7}_3$7$_5$  & ({\bf 1, 1}) & - 1 & 0 & 0 & 0 & 1 \\
\hline
 (0, -1, 0) & 7$_5$${\tilde 7}_3$  & ({\bf 1, 1})&   0 & 0 & 1 & 0 & -1 \\
\hline
\end{tabular}
\end{center}

\subsection{Gauge groups and chiral fermion spectra at fixed points
($0, 1, \pm 1 $) and ($0, -1, \pm 1$)}

The twisted R-R tadpole   cancellation conditions at these four  fixed
points have been given in Eqs.(\ref{eq: 29}).  With the assignments of
Eq.(\ref{eq:  36}),   we   get  the following   solutions   for  these
conditions:
\begin{equation} \label{eq: 42}
\left.
\begin{array}{lllll}
\gamma_{1, 3, (0, 1, \pm 1)} &  = &  \gamma_{1, 3, (0, -1, \pm 1)} &
= &  {\bf 0}  \\
\gamma_{1, {\tilde 7}_{3}, (0)}\gamma^{\pm 1}_{W4, {\tilde 7}_3} &  = &  - \alpha {\bf
1}_{1} &    &  \\
\gamma_{1, 7_{4}, (\pm 1)} &  =  &  \alpha {\bf 1}_{1} &   & \\
\gamma_{1, 7_{5}, (\pm 1)} &  =  &  {\bf 0} &   &
\end{array}
\right.
\end{equation}
These orbifold  actions  on Chan-Paton wavefunctions imply  that there
are no D3- and D7$_{5}$-branes located at these fixed points. In other
words,  these fixed points  are  only wrapped by  ${\widetilde{\textrm
D7}}_{3,(0)}$-branes and D7$_{4, (\pm 1)}$-branes with which the gauge
groups  $U(1)_{1,   (0)}$  and  $U(1)_{2,  (\pm   1)}$  are associated
respectively. The Chan-Paton  wavefunctions at these fixed points  are
determined by,
\begin{equation} \label{eq: 43}
\left.
\begin{array}{ll}
\widetilde{\textrm{D7}}_{3}\textrm{D7}_{4}: \,\,\,\, & \lambda \, = \, \alpha
\, [\, \gamma_{1, {\tilde 7}_3, (0)} \gamma_{W4, {\tilde 7}_3}^{\pm 1}\, ] \, \lambda
\, \gamma^{-1}_{1, 7_4, (\pm 1)} \\
\textrm{D7}_{4}\widetilde{\textrm{D7}}_{3}: \,\,\,\, & \lambda \, = \,\, \alpha \,
\gamma_{1, 7_4, (\pm  1)}  \, \lambda \, \, [ \,  \gamma_{1, {\tilde 7}_3, (0)}
\gamma_{W4, {\tilde 7}_3}^{\pm 1} \, ]^{-1}
\end{array}
\right.
\end{equation}
There is no qualified massless chiral fermionic state subject to these
constraints.

\subsection{Gauge groups and chiral fermion spectra at fixed points
($m, 0,  \pm 1 $) }

The  twisted R-R tadpole  cancellation  conditions at these six  fixed
points have been given in Eqs.(\ref{eq: 31}).  With the assignments of
Eq.(\ref{eq: 36}),   we     get the following    solutions   for these
conditions:
\begin{equation} \label{eq: 44}
\left.
\begin{array}{lll}
\gamma_{1, 3, (m, 0, \pm 1)} & = & - \alpha {\bf 1}_{1}  \\
\gamma_{1, {\tilde 7}_{3}, (m)} &  = &  - \alpha {\bf 1}_{1} \\
\gamma_{1, 7_{4}, (0)} &  = &  - \alpha {\bf 1}_{2} \\
\gamma_{1, 7_{5}, (\pm 1)} &  =  &  {\bf 0}
\end{array}
\right.
\end{equation}
Therefore, there is one D3-brane at  each (considered) fixed point. In
addition,    each  of  these    six     points   is  wrapped by      a
${\widetilde{\textrm D7}}_{3}$-brane and  two D7$_{4}$-branes.  On the
worldvolumes of  these D-branes and  anti-branes  the gauge groups are
distributed as,
\begin{equation} \label{eq: 45}
\left.
\begin{array}{ll}
\textrm{D3$_{(m, 0, \pm 1)}$}: &  U(1)_{0, (m, 0, \pm 1)} \\
\widetilde{\textrm{D7}}_{3, (m)}: &  U(1)_{1, ( m )}  \\
\textrm{D7$_{4,  ( 0 )}$}: &  U(2) = U(1)_{2, (0)} \times SU(2)
\end{array}
\right.
\end{equation}
There is no massless left-handed fermionic state created from the open
string sectors near these fixed points.

\subsection{Gauge groups and chiral fermion spectra at fixed points
($\pm 1, 1, 0 $) and ($\pm 1, -1, 0$)}

The twisted R-R tadpole   cancellation conditions at these four  fixed
points have been given in Eqs.(\ref{eq: 32}).  With the assignments of
Eq.(\ref{eq:  36}),   we   get  the following   solutions   for  these
conditions:
\begin{equation} \label{eq: 46}
\left.
\begin{array}{lllll}
\gamma_{1, 3, (\pm 1, 1, 0)} &  = &  \gamma_{1, 3, (\pm 1, -1, 0)}
 \,\,=  \,{\bf 0} &  &\\
\gamma_{1, {\tilde 7}_{3}, (\pm 1)}   &  = &  - \alpha {\bf 1}_{1} &    &  \\
\gamma_{1, 7_{4}, (\pm 1)} &  =  &  \alpha {\bf 1}_{1} &   & \\
\gamma_{1, 7_{5}, (0)} \gamma_{W4, 7_5} &  =  & \textrm{diag}(\alpha^2
{\bf 1}_{1}, {\bf 1}_{1}, \alpha {\bf 1}_{1} ) &   & \\
\gamma_{1, 7_{5}, (0)} \gamma^{-1}_{W4, 7_5} &  =  & \textrm{diag}(\alpha^2
{\bf 1}_{1},  \alpha {\bf 1}_{1}, {\bf 1}_{1} ) &   &
\end{array}
\right.
\end{equation}
Thereby, every    fixed  point in    this   set is  wrapped    by  one
${\widetilde{\textrm D7}}_{3}$-brane,   one D7$_{4}$-brane  and  three
D7$_{5}$-branes. There is no D3-branes located at these points. Because
of  the presence of  the Wilson  line passing  through the origin  and
along  the $4$-th  complex plane,  the gauge  groups near  these fixed
points  at  respectively $U(1)_{1, (\pm   1)}$ on ${\widetilde{\textrm
D7}}_{3,  (\pm  1)}$-branes,   $U(1)_{2, (\pm  1)}$  on  D7$_{4,  (\pm
1)}$-branes  and   $U(1)_{31}  \times U(1)_{32} \times  U(1)_{33}$  on
D7$_{5, (0)}$-branes.  The qualified Chan-Paton wavefunctions at these
fixed points satisfy constraint conditions:
\begin{equation} \label{eq: 47}
\left.
\begin{array}{lllll}
\widetilde{\textrm{D7}}_{3}\textrm{D7}_{4}: &\,\,\,\,\,  \lambda & = & \alpha
 \gamma_{1, {\tilde 7}_{3}, (\pm 1)}  \lambda \gamma^{-1}_{1, 7_4, (n)}, & (n =
 \pm 1) \\
\textrm{D7}_{4}\widetilde{\textrm{D7}}_{3}: &\,\,\,\,\,  \lambda & = & \alpha
\gamma_{1, 7_4, (n)}  \lambda    \gamma^{-1}_{1, {\tilde 7}_{3}, (\pm 1)}, & (n
 = \pm 1) \\
\widetilde{\textrm{D7}}_{3}\textrm{D7}_{5}: &\,\,\,\,\,  \lambda & = & - \alpha
 \gamma_{1, {\tilde 7}_{3}, (\pm 1)}  \lambda [\gamma_{1, 7_{5}, (0)}
 \gamma^{n}_{W4, 7_5}]^{-1}, & (n = \pm 1 ) \\
\textrm{D7}_{5}\widetilde{\textrm{D7}}_{3}: & \,\,\,\,\, \lambda & = & - \alpha
 [\gamma_{1, 7_{5}, (0)}\gamma^{n}_{W4, 7_5}] \lambda \gamma^{-1}_{1,
 {\tilde 7}_{3}, (\pm 1)}, & ( n = \pm 1) \\
\textrm{D7}_{4}\textrm{D7}_{5}: &\,\,\,\,\,  \lambda & = & - \alpha^2
 \gamma_{1, 7_{4}, (n)}  \lambda [\gamma_{1, 7_{5}, (0)}
 \gamma^{n}_{W4, 7_5}]^{-1}, & (n = \pm 1 ) \\
\textrm{D7}_{5}\textrm{D7}_{4}: & \,\,\,\,\, \lambda & = & - \alpha^2
 [\gamma_{1, 7_{5}, (0)}\gamma^{n}_{W4, 7_5}] \lambda \gamma^{-1}_{1,
 7_{4}, (n)}, & ( n  = \pm 1 )
\end{array}
\right.
\end{equation}
There are   totally  eight  massless left-handed   fermion  states
created from the open string sectors near these four fixed points (two
states    for  each points).    These  chiral   states form   the unit
representations  for  both  $SU(3)$  and  $SU(2)$.  We summarize their
$U(1)$-charges in the following table:
\begin{center}
\begin{tabular}{|r|r|r|r|r|r|c|c|c|} \hline
\multicolumn{9}{|c|}{\bfseries{Massless Chiral Fermion Spectra at
($\pm 1, 1, 0$) and ($\pm 1, -1, 0$) }} \\ \hline
 Fixed Point & Sector  & State & Q$_{31}$ & Q$_{32}$ & Q$_{33}$ & Q$_{2,
 (\pm 1)}$ & Q$_{1, ( 1)}$ & Q$_{1, (-1)}$ \\
\hline
(1, 1, 0) & ${\tilde 7}_3$7$_5$  & ({\bf 1, 1}) & - 1 & 0 & 0 & 0 & 1 & 0 \\
\hline
(1, 1, 0) & 7$_5$${\tilde 7}_3$  & ({\bf 1, 1}) & 0   & 1 & 0 & 0 & -1 & 0 \\
\hline
(-1, 1, 0) & ${\tilde 7}_3$7$_5$  & ({\bf 1, 1})& - 1 & 0 & 0 & 0 & 0 & 1 \\
\hline
(-1, 1, 0) & 7$_5$${\tilde 7}_3$  & ({\bf 1, 1}) & 0   & 1 & 0 & 0 & 0  & -1 \\
\hline
(1, -1, 0) & ${\tilde 7}_3$7$_5$  & ({\bf 1, 1}) & - 1 & 0 & 0 & 0 & 1 & 0 \\
\hline
(1, -1, 0) & 7$_5$${\tilde 7}_3$  & ({\bf 1, 1}) & 0   & 0 & 1 & 0 & -1 & 0\\
\hline
(-1, -1, 0) & ${\tilde 7}_3$7$_5$  & ({\bf 1, 1})& - 1 & 0 & 0 & 0 & 0 & 1 \\
\hline
(-1, -1, 0) & 7$_5$${\tilde 7}_3$  & ({\bf 1, 1}) & 0   & 0 & 1 & 0 & 0 & -1\\
\hline
\end{tabular}
\end{center}

\subsection{Gauge groups and chiral fermion spectra at fixed points
($m, n, p$) with $m, n, p = \pm 1$}

The twisted R-R tadpole   cancellation conditions at these eight  fixed
points have been listed in Eqs.(\ref{eq: 34}).  With the assignments of
Eq.(\ref{eq:  36}),   we have  the following solutions:
\begin{equation} \label{eq: 48}
\left.
\begin{array}{llrl}
\gamma_{1, 3, (m, n, p)} &  = &  {\bf 0}, & (m, n, p = \pm 1)  \\
\gamma_{1, {\tilde 7}_{3}, (m)}   &  = &  - \alpha {\bf 1}_{1}, & (m = \pm 1)\\
\gamma_{1, 7_{4}, (n)}   &  = &  \alpha {\bf 1}_{1},   & (n = \pm 1)\\
\gamma_{1, 7_{5}, (p)}   &  = &  {\bf 0}, & (p = \pm 1)
\end{array}
\right.
\end{equation}
Therfore, there is no D3-branes located at these eight fixed points. In
fact, each of    these  points is supposed  to   be  wrapped   by  one
${\widetilde{\textrm  D7}}_{3}$-brane  and   one D7$_{4}$-brane  only.
There  is an abelian gauge  group $U(1)_{1, {m}} \times U(1)_{2, (n)}$
on the D-brane worldvolume  near each fixed point  $(m, n, p)$ in this
set. However,  there is no  massless left-handed fermion state created
from the corresponding open string sectors.

Let  us make   a brief   summary for  the   obtained massless  fermion
spectra. At fixed points $(m, 0, 0)$ with $m=0, \pm 1$, we get exactly
the  non-supersymmetric Standard   model chiral fermion  spectra  with
{\it three} generations. At the other six fixed points $(m, \pm 1,
0)$ with  $m = 0,  \pm1$, there are  twelve extra massless left-handed
fermions which nevertheless form   the  singlets of  nonabelian  gauge
group  $SU(3) \times SU(2)$.  There  are no  massless chiral  fermions
located at remaining eighteen fixed points.

\section{Gauge Anomalies and Cancellation Mechanism}
\renewcommand{\theequation}{4.\arabic{equation}}\setcounter{equation}{0}

According  to the obtained  chiral fermion  spectra,  there are no any
cubic nonabelian gauge anomalies  at  all ${\bf Z}_3$ orbifold   fixed
points. There are also not any  $U(1)$ anomalies at these fixed points
except at $(0, 0,  0)$  and $(\pm 1 ,  0,  0)$. There exist  potential
$U(1)$   anomalies    at these three  fixed   points.   Relying on the
coincidence  of  the  massless chiral  fermionic  spectra,  it is only
necessary to discuss the $U(1)$ anomalies at orbifold origin and their
cancellation mechanism.

\subsection{Mixed $U(1)$-nonabelian anomalies}

Taking   the following normalization   for $SU(n)$'s generators (where
$n=2,3$),
\begin{equation} \label{eq: 49}
\left. \begin{array}{ll}
Tr(\lambda_a \lambda_b)= \frac{1}{2} \delta_{ab}, & \,\,\, Tr(\lambda_a)= 0
\end{array}
\right.
\end{equation}
we can easily get from the massless left-handed fermion spectrum at
origin that,
\begin{equation} \label{eq: 50}
\left. \begin{array}{lcrclcr}
Tr\,[\,Q_{0} \lambda^{2}_{SU(2)} \,]\,&  = & -1 & \, \, \, &
Tr\,[\,Q_{0} \lambda^{2}_{SU(3)} \,]\,&  = &  1 \\
Tr\,[\,Q_{1} \lambda^{2}_{SU(2)} \,]\,&  = &  0 & \, \, \, &
Tr\,[\,Q_{1} \lambda^{2}_{SU(3)} \,]\,&  = &  1 \\
Tr\,[\,Q_{2} \lambda^{2}_{SU(2)} \,]\,&  = & -2 & \, \, \, &
Tr\,[\,Q_{2} \lambda^{2}_{SU(3)} \,]\,&  = & -2 \\
Tr\,[\,Q_{3} \lambda^{2}_{SU(2)} \,]\,&  = &  3 & \, \, \, &
Tr\,[\,Q_{3} \lambda^{2}_{SU(3)} \,]\,&  = &  0
\end{array}
\right.
\end{equation}
For sake of convenience, we have simply denoted the $U(1)$ charges
$\{ Q_{0, {(0,0,0)}}, \, Q_{1, (0)} \}$ as $\{ Q_{0}, \, Q_{1} \}$ in
Eq.(\ref{eq: 50})  (    and thereafter).  These  traces  describe  the
possible mixed $U(1)$-nonabelian anomalies at the orbiford origin.  If
we  introduce  alternatively  another set  of   the orthogonal  $U(1)$
charges
\begin{equation} \label{eq: 51}
\left \{
\begin{array}{l}
Y = \frac{2}{3} Q_3 + \frac{1}{2} Q_2 + Q_{0} \\
\widetilde{Q_{1}} = Q_{1} \\
\widetilde{Q_{2}} = 3 a Q_{3} + 2 b Q_{2} - (2a + b) Q_{0} \\
\widetilde{Q_{3}} = 3 c Q_{3} + 2 d Q_{2} - (2c + d) Q_{0}
\end{array}
\right.
\end{equation}
where,
\begin{equation} \label{eq: 52}
\left. \begin{array}{llllr}
a = \sqrt{21533} + 119 \, & \,\,\,& b & = &  - 194 \\
c = \sqrt{21533} - 119 \, & \,\,\,& d & = &  194
\end{array}
\right.
\end{equation}
and interpret $Y$ as the weak hypercharge at  the origin, we can prove
that,   among these four    independent   $U(1)$  charges,  only   the
hypercharge is free of the $U(1)$-nonabelian gauge anomalies:
\begin{equation} \label{eq: 53}
Tr\,[\,Y  \lambda^{2}_{SU(2)}\,]\,   =     Tr\,[\,Y\lambda^{2}_{SU(3)}
\,]\,=\,0
\end{equation}
The other three ones remain anomalous.

\subsection{Pure $U(1)$ anomalies}

The pure $U(1)$ anomalies  can  be categorized  into three types:  the
cubic $U(1)$ anomalies,  the  mixed $U(1)$  anomalies and the   triple
mixed $U(1)$  anomalies. At  ${\bf Z_3}$   orbifold origin, the  cubic
$U(1)$ anomalies are measured by traces:
\begin{equation} \label{eq: 54}
\left.   \begin{array}{llll}  Tr\,[\,Q^{3}_{0}\,]\,=    4   \,  &   \,
Tr\,[\,Q^{3}_{1}\,]\,= 4  \, & \,  Tr\,[\,Q^{3}_{2}\,]\, = -8 \,  & \,
Tr\,[\,Q^{3}_{3}\,]\, = 0
\end{array}
\right.
\end{equation}
The mixed $U(1)$ anomalies are measured by the following traces:
\begin{equation} \label{eq: 55}
\left. \begin{array}{llllr}
Tr\,[\,Q^{2}_{0} Q_{1}\,]\, & = & - \,
Tr\,[\,Q^{2}_{1} Q_{0}\,]\, & = & -2 \\
Tr\,[\,Q^{2}_{0} Q_{2}\,]\, & = & - \,
Tr\,[\,Q^{2}_{2} Q_{0}\,]\, & = & 4 \\
Tr\,[\,Q^{2}_{0} Q_{3}\,]\, & = & - \,
Tr\,[\,Q^{2}_{3} Q_{0}\,]\, & = & -6 \\
Tr\,[\,Q^{2}_{1} Q_{2}\,]\, & = & - \,
Tr\,[\,Q^{2}_{2} Q_{1}\,]\, & = & 0 \\
Tr\,[\,Q^{2}_{1} Q_{3}\,]\, & = & - \,
Tr\,[\,Q^{2}_{3} Q_{1}\,]\, & = & -6 \\
Tr\,[\,Q^{2}_{2} Q_{3}\,]\, & = & - \,
Tr\,[\,Q^{2}_{3} Q_{2}\,]\, & = & 12
\end{array}
\right.
\end{equation}
In the  obtained chiral fermionic  spectrum, each  left-handed fermion
state form the fundamental representation of  one factoral gauge group
and   the anti-fundamental representation  of  another  factoral gauge
group.  There is no such a  fermion state that has three non-vanishing
distinct $U(1)$ charges.  Therefore, the model  under consideration is
free of the triple mixed $U(1)$ anomalies:
\begin{equation} \label{eq: 57}
\left.  \begin{array}{l}
Tr[Q_{0} Q_{1} Q_{2}] = 0 \\
Tr[Q_{0} Q_{1} Q_{3}] = 0 \\
Tr[Q_{0} Q_{2} Q_{3}] = 0 \\
Tr[Q_{1} Q_{2} Q_{3}] = 0
\end{array}
\right.
\end{equation}
Particularly, the hypercharge $Y$ is free of the cubic $U(1)$ anomaly:
\begin{equation} \label{eq: 58}
Tr(Y^{3}) =0
\end{equation}

\subsection{Factorized anomaly traces and mechanism for $U(1)$ anomaly
cancellation}

It is remarkable that all the $U(1)$ anomalies  appearing in our model
can           be          cancelled        through       Green-Schwarz
mechanism\cite{9808139}\cite{9904071}.   To      make  this conclusion
transparent,  we   now  search  the   equivalent  but  more insightful
expressions of the above  anomaly traces\cite{9808139}.  First, let us
define an extended Delta symbol in ${\bf Z}_{3}$ orbifold case,
\begin{equation} \label{eq: 59}
\delta_{jl} = \left \{
\begin{array}{llll}
1 & \textrm{if} \,\,\, j-l= 3N & (N \textrm{ is an arbitrary integer}) \\
0 & \textrm{otherwise}  &
\end{array}
\right.
\end{equation}
Then,
\begin{equation} \label{eq: 60}
\left.
\begin{array}{llr}
\sum_{r=3}^{5} ( \delta_{2, 0 + b_r } - \delta_{2, 0 - b_r } ) & = &
-3 \\
\sum_{r=3}^{5} ( \delta_{2, 3 + b_r } - \delta_{2, 3 - b_r } ) & = &
-3 \\
\sum_{r=3}^{5} ( \delta_{3, 1 + b_r } - \delta_{3, 1 - b_r } ) & = &
-3 \\
\sum_{r=3}^{5} ( \delta_{3, 2 + b_r } - \delta_{3, 2 - b_r } ) & = &
3
\end{array}
\right.
\end{equation}
With the help of the discrete Fourier transformation
\begin{equation} \label{eq: 61}
3 \delta_{l, \, j \pm b_r} = \sum_{k=1}^{3} \exp \,[\,2 \pi i k (j - l
\pm b_r)/3\,]
\end{equation}
and the mathematical identity
\begin{equation} \label{eq: 62}
\sum_{r=3}^{5} \sin(\frac{2 \pi k b_r}{3}) = - \frac{1}{2} \prod^{5}_{r=3}
\,[\,2 \sin(\frac{\pi k b_r}{3}) \,]
\end{equation}
we can obtain from Eqs.(\ref{eq: 60}) the following two significant
mathematics formulae:
\begin{equation} \label{eq: 63}
1 = \, \pm \frac{i}{9} \sum_{k=1}^{3} \,[\,\prod_{r=3}^{5} 2
\sin(\frac{\pi k b_r}{3}) \,]\,\exp(\pm \frac{2 \pi i k}{3})
\end{equation}

On the other hand, it follows directly from Eq.(\ref{eq: 4}) that,
\begin{equation} \label{eq: 64}
\left.
\begin{array}{rll}
1 & = & Tr\gamma_{k, 3, (0, 0, 0)} \\
\exp[\,{(2 \pi i k)}/{3}\,] & = & (-1)^{k}Tr\gamma_{k, {\tilde 7}_3, (0)} \\
2\exp[\,{(2 \pi i k)}/{3}\,] & = & (-1)^{k}Tr\gamma_{k, 7_4, (0)} \\
3\exp[\,- {(2 \pi i k)}/{3}\,] & = & Tr\gamma_{k, 7_5, (0)}
\end{array}
\right.
\end{equation}
and
\begin{equation} \label{eq: 65}
\left.
\begin{array}{rll}
1 & = & Tr\gamma_{k, 3, (0, 0, 0)}^{-1} \\
\exp[\,- {(2 \pi i k)}/{3}\,] & = & (-1)^{k}Tr\gamma_{k, {\tilde 7}_3, (0)}^{-1}
\\
2\exp[\,- {(2 \pi i k)}/{3}\,] & = & (-1)^{k}Tr\gamma_{k, 7_4, (0)}^{-1}
\\
3\exp[\,  {(2 \pi i k)}/{3}\,] & = & Tr\gamma_{k, 7_5, (0)}^{-1}
\end{array}
\right.
\end{equation}
where $k=1, 2, 3$. Therfore, the cubic $U(1)$ anomaly traces in
Eq.(\ref{eq: 54}) can be rewritten as,
\begin{equation} \label{eq: 66}
\left. \begin{array}{lll}
\frac{1}{3} Tr [\,Q^{3}_{0}\,]\, &  = & \frac{4}{3} \\
 & = & \frac{4}{3} \cdot \frac{1}{3} \sum^{3}_{k=1} Tr(\gamma_{k, 3, (0,
0, 0)}) Tr(\gamma_{k, 3, (0, 0, 0)}^{-1}) \\
 & = & \frac{4}{9} \sum^{3}_{k=1} Tr[\, \gamma_{k, 3, (0,0, 0)}
\lambda_{U(1)_{0, (0, 0, 0)}}\,]\, Tr[\,\gamma_{k, 3, (0, 0, 0)}^{-1}
\lambda^{2}_{U(1)_{0, (0, 0, 0)}}\,]
\end{array}
\right.
\end{equation}
\begin{equation} \label{eq: 67}
\left. \begin{array}{lll}
\frac{1}{3}   Tr [\,Q^{3}_{1}\,]\,   & =  &   \frac{4}{3} \\    &  = &
 \frac{4}{3}  \cdot  \frac{1}{3}  \sum^{3}_{k=1} Tr(\gamma_{k, {\tilde
 7}_3,   (0)})  Tr(\gamma_{k,  {\tilde  7}_3,  (0)}^{-1})   \\ &  =  &
 \frac{4}{9} \sum^{3}_{k=1}  Tr[\,    \gamma_{k,  {\tilde 7}_3,   (0)}
 \lambda_{U(1)_{1, (0)}}\,]\, Tr[\,\gamma_{k, {\tilde 7}_3,  (0)}^{-1}
 \lambda^{2}_{U(1)_{1, (0)}}\,]
\end{array}
\right.
\end{equation}
\begin{equation} \label{eq: 68}
\left. \begin{array}{lll}
\frac{1}{3} Tr [\,Q^{3}_{2}\,]\, &  = & - \frac{8}{3} \\
 & = & - \frac{8}{3} \cdot \frac{1}{12} \sum^{3}_{k=1} Tr(\gamma_{k, 7_4,
(0)}) Tr(\gamma_{k, 7_4, (0)}^{-1}) \\
 & = & - \frac{2}{9} \sum^{3}_{k=1} Tr[\, \gamma_{k, 7_4, (0)}
\lambda_{U(1)_{2}}\,]\, Tr[\,\gamma_{k, 7_4, (0)}^{-1}
\lambda^{2}_{U(1)_{2}}\,]
\end{array}
\right.
\end{equation}
and $
\frac{1}{3} Tr [\,Q^{3}_{3}\,]\, =  0
$
. The mixed $U(1)$ anomaly traces can also acquire similar expressions:
\begin{equation} \label{eq: 69}
\left \{  \begin{array}{lll}
Tr[\,Q_{0}^{2}Q_{1}\,]
      & = & - \frac{2i}{9} \sum_{k=1}^{3} (-1)^{k}
             \, [\,\prod_{r=3}^{5} 2 \sin({\pi k b_r}/{3})\,]\,\\
      &   & \,\,\, \cdot Tr[\,\gamma_{k, {\tilde 7}_3, (0)} \lambda_{U(1)_{1,
              (0)}}\,]\, Tr[\,\gamma^{-1}_{k, 3, (0, 0, 0)}
              \lambda^{2}_{U(1)_{0, (0, 0, 0)}}\,] \\
Tr[\,Q_{0}^{2}Q_{2}\,]
      & = &   \frac{2i}{9} \sum_{k=1}^{3} (-1)^{k}
             \, [\,\prod_{r=3}^{5} 2 \sin({\pi k b_r}/{3})\,]\,\\
      &   & \,\,\, \cdot Tr[\,\gamma_{k, 7_4, (0)} \lambda_{U(1)_{2, (0)
               }}\,]\, Tr[\,\gamma^{-1}_{k, 3, (0, 0, 0)}
              \lambda^{2}_{U(1)_{0, (0, 0, 0)}}\,] \\
Tr[\,Q_{0}^{2}Q_{3}\,]
      & = &   \frac{2i}{9} \sum_{k=1}^{3}
             \, [\,\prod_{r=3}^{5} 2 \sin({\pi k b_r}/{3})\,]\,\\
      &   & \,\,\, \cdot Tr[\,\gamma_{k, 7_5, (0)} \lambda_{U(1)_{3
               }}\,]\, Tr[\,\gamma^{-1}_{k, 3, (0, 0, 0)}
              \lambda^{2}_{U(1)_{0, (0, 0, 0)}}\,]
\end{array}
\right.
\end{equation}
\begin{equation} \label{eq: 70}
\left \{  \begin{array}{lll}
Tr[\,Q_{1}^{2}Q_{0}\,]
      & = & - \frac{2i}{9} \sum_{k=1}^{3} (-1)^{k}
             \, [\,\prod_{r=3}^{5} 2 \sin({\pi k b_r}/{3})\,]\,\\
      &   & \,\,\, \cdot Tr[\,\gamma_{k, 3, (0, 0, 0)} \lambda_{U(1)_{0,
              (0, 0, 0)}}\,]\, Tr[\,\gamma^{-1}_{k, {\tilde 7}_3, (0)}
              \lambda^{2}_{U(1)_{1, (0)}}\,] \\
Tr[\,Q_{1}^{2}Q_{2}\,] & = &  0  \\
Tr[\,Q_{1}^{2}Q_{3}\,]
      & = & - \frac{2i}{9} \sum_{k=1}^{3} (-1)^{k}
             \, [\,\prod_{r=3}^{5} 2 \sin({\pi k b_r}/{3})\,]\,\\
      &   & \,\,\, \cdot Tr[\,\gamma_{k, 7_5, (0)} \lambda_{U(1)_{3
               }}\,]\, Tr[\,\gamma^{-1}_{k, {\tilde 7}_3, (0)}
              \lambda^{2}_{U(1)_{1, (0)}}\,]
\end{array}
\right.
\end{equation}
\begin{equation} \label{eq: 71}
\left \{  \begin{array}{lll}
Tr[\,Q_{2}^{2}Q_{0}\,]
      & = &  \frac{2i}{9} \sum_{k=1}^{3} (-1)^{k}
             \, [\,\prod_{r=3}^{5} 2 \sin({\pi k b_r}/{3})\,]\,\\
      &   & \,\,\, \cdot Tr[\,\gamma_{k, 3, (0, 0, 0)} \lambda_{U(1)_{0,
              (0, 0, 0)}}\,]\, Tr[\,\gamma^{-1}_{k, 7_4, (0)}
              \lambda^{2}_{U(1)_{2, (0)}}\,] \\
Tr[\,Q_{2}^{2}Q_{1}\,] & = &  0  \\
Tr[\,Q_{2}^{2}Q_{3}\,]
      & = &  \frac{2i}{9} \sum_{k=1}^{3} (-1)^{k}
             \, [\,\prod_{r=3}^{5} 2 \sin({\pi k b_r}/{3})\,]\,\\
      &   & \,\,\, \cdot Tr[\,\gamma_{k, 7_5, (0)} \lambda_{U(1)_{3
               }}\,]\, Tr[\,\gamma^{-1}_{k, 7_4, (0)}
              \lambda^{2}_{U(1)_{2, (0)}}\,]
\end{array}
\right.
\end{equation}
\begin{equation} \label{eq: 72}
\left \{  \begin{array}{lll}
Tr[\,Q_{3}^{2}Q_{0}\,]
      & = &  \frac{2i}{9} \sum_{k=1}^{3}
             \, [\,\prod_{r=3}^{5} 2 \sin({\pi k b_r}/{3})\,]\,\\
      &   & \,\,\, \cdot Tr[\,\gamma_{k, 3, (0, 0, 0)} \lambda_{U(1)_{0,
              (0, 0, 0)}}\,]\, Tr[\,\gamma^{-1}_{k, 7_5, (0)}
              \lambda^{2}_{U(1)_{3}}\,] \\
Tr[\,Q_{3}^{2}Q_{1}\,]
      & = &  -  \frac{2i}{9} \sum_{k=1}^{3} (-1)^{k}
             \, [\,\prod_{r=3}^{5} 2 \sin({\pi k b_r}/{3})\,]\,\\
      &   & \,\,\, \cdot Tr[\,\gamma_{k, {\tilde 7}_3, (0)} \lambda_{U(1)_{1, (0)
               }}\,]\, Tr[\,\gamma^{-1}_{k, 7_5, (0)}
              \lambda^{2}_{U(1)_{3}}\,] \\
Tr[\,Q_{3}^{2}Q_{2}\,]
      & = &   \frac{2i}{9} \sum_{k=1}^{3} (-1)^{k}
             \, [\,\prod_{r=3}^{5} 2 \sin({\pi k b_r}/{3})\,]\,\\
      &   & \,\,\, \cdot Tr[\,\gamma_{k, 7_4, (0)} \lambda_{U(1)_{2, (0)
               }}\,]\, Tr[\,\gamma^{-1}_{k, 7_5, (0)}
              \lambda^{2}_{U(1)_{3}}\,]
\end{array}
\right.
\end{equation}

The traces (\ref{eq: 50}) measuring mixed $U(1)$ nonabelian anomalies
can  also be expressed  in the similar (factorized) fashion. Noticing,
$Tr[\,\lambda^{2}_{SU(2)}\,]=Tr[\,       \lambda^{2}_{SU(3)}\,]      =
\frac{1}{2}$, we have:
\begin{equation} \label{eq: 73}
\left. \begin{array}{lcl}
Tr[\,\gamma^{-1}_{k, 7_4}\lambda^{2}_{SU(2)}\,]\,=\frac{1}{2}(-1)^{k}
\exp(-{2 \pi i k}/{3}) & \,&
Tr[\,\gamma^{-1}_{k, 7_5}\lambda^{2}_{SU(3)}\,]\,=\frac{1}{2}
\exp({2 \pi i k}/{3})
\end{array}
\right.
\end{equation}
Moreover,
%
$$
\left. \begin{array}{lll}
Tr[\,\gamma_{k, 3, (0, 0, 0)} \lambda_{U(1)_{0, (0, 0, 0)}}\,]\,
Tr[\,\gamma^{-1}_{k, 7_4}\lambda^{2}_{SU(2)}\,]\, & = & \frac{1}{2}(-1)^{k}
\exp(-{2 \pi i k}/{3}) \\
Tr[\,\gamma_{k, 3, (0, 0, 0)} \lambda_{U(1)_{0, (0, 0, 0)}}\,]\,
Tr[\,\gamma^{-1}_{k, 7_5}\lambda^{2}_{SU(3)}\,]\, & = & \frac{1}{2}
\exp({2 \pi i k}/{3}) \\
Tr[\,\gamma_{k, {\tilde 7}_3, (0)} \lambda_{U(1)_{1, (0)}}\,]\,
Tr[\,\gamma^{-1}_{k, 7_5}\lambda^{2}_{SU(3)}\,]\, & = & \frac{1}{2}(-1)^{k}
\exp(-{2 \pi i k}/{3}) \\
Tr[\,\gamma_{k, 7_4, (0)} \lambda_{U(1)_{2, (0)}}\,]\,
Tr[\,\gamma^{-1}_{k, 7_4}\lambda^{2}_{SU(2)}\,]\, & = & 1 \\
Tr[\,\gamma_{k, 7_4, (0)} \lambda_{U(1)_{2, (0)}}\,]\,
Tr[\,\gamma^{-1}_{k, 7_5}\lambda^{2}_{SU(3)}\,]\, & = & (-1)^{k}
\exp(-{2 \pi i k}/{3}) \\
Tr[\,\gamma_{k, 7_5, (0)} \lambda_{U(1)_{3}}\,]\,
Tr[\,\gamma^{-1}_{k, 7_4}\lambda^{2}_{SU(2)}\,]\, & = & \frac{3}{2}
(-1)^{k}\exp({2 \pi i k}/{3})
\end{array}
\right.
$$
%
Rewriting the non-vanishing traces in Eqs.(\ref{eq: 50}) with these
mathematical identities, we get:
\begin{equation} \label{eq: 75}
\left.  \begin{array}{lll}
Tr[\,Q_{0} \lambda^{2}_{SU(2)}\,]
      & = &  \frac{2i}{9} \sum_{k=1}^{3} (-1)^{k}
             \, [\,\prod_{r=3}^{5} 2 \sin({\pi k b_r}/{3})\,]\,\\
      &   & \,\,\, \cdot Tr[\,\gamma_{k, 3, (0, 0, 0)} \lambda_{U(1)_{0,
              (0, 0, 0)}}\,]\, Tr[\,\gamma^{-1}_{k, 7_4, (0)}
              \lambda^{2}_{SU(2)}\,] \\
Tr[\,Q_{0} \lambda^{2}_{SU(3)}\,]
      & = &    \frac{2i}{9} \sum_{k=1}^{3}
             \, [\,\prod_{r=3}^{5} 2 \sin({\pi k b_r}/{3})\,]\,\\
      &   & \,\,\, \cdot Tr[\,\gamma_{k, 3, (0, 0, 0)}\lambda_{U(1)_{0, (0, 0, 0)
               }}\,]\, Tr[\,\gamma^{-1}_{k, 7_5, (0)}
              \lambda^{2}_{SU(3)}\,] \\
Tr[\,Q_{1} \lambda^{2}_{SU(3)}\,]
      & = & -  \frac{2i}{9} \sum_{k=1}^{3} (-1)^{k}
             \, [\,\prod_{r=3}^{5} 2 \sin({\pi k b_r}/{3})\,]\,\\
      &   & \,\,\, \cdot Tr[\,\gamma_{k, {\tilde 7}_3, (0)} \lambda_{U(1)_{1, (0)
               }}\,]\, Tr[\,\gamma^{-1}_{k, 7_5, (0)}
              \lambda^{2}_{SU(3)}\,]\\
Tr[\,Q_{2} \lambda^{2}_{SU(3)}\,]
      & = &   \frac{2i}{9} \sum_{k=1}^{3} (-1)^{k}
             \, [\,\prod_{r=3}^{5} 2 \sin({\pi k b_r}/{3})\,]\,\\
      &   & \,\,\, \cdot Tr[\,\gamma_{k, 7_4, (0)} \lambda_{U(1)_{2, (0)
               }}\,]\, Tr[\,\gamma^{-1}_{k, 7_5, (0)}
              \lambda^{2}_{SU(3)}\,]\\
Tr[\,Q_{3} \lambda^{2}_{SU(2)}\,]
      & = &   \frac{2i}{9} \sum_{k=1}^{3} (-1)^{k}
             \, [\,\prod_{r=3}^{5} 2 \sin({\pi k b_r}/{3})\,]\,\\
      &   & \,\,\, \cdot Tr[\,\gamma_{k, 7_5, (0)} \lambda_{U(1)_{3
               }}\,]\, Tr[\,\gamma^{-1}_{k, 7_4, (0)}
              \lambda^{2}_{SU(2)}\,]
\end{array}
\right.
\end{equation}
and
\begin{equation} \label{eq: 76}
Tr[\,Q_{2} \lambda^{2}_{SU(2)}\,]
      =  -  \frac{2}{3} \sum_{k=1}^{3} Tr[\,\gamma_{k, 7_4, (0)}
            \lambda_{U(1)_{2, (0) }}\,]\, Tr[\,\gamma^{-1}_{k, 7_4, (0)}
            \lambda^{2}_{SU(2)}\,]
\end{equation}

In view of the above factorized expressions of the $U(1)$ anomaly
traces, we easily see that they can be cancelled by the coupling in
the closed string sectors of the $U(1)$ to a R-R field $B^{\mu
\nu}_{k}$,
$$
Tr(\gamma_{k} \lambda_{i}) B_{k} \wedge F_{U(1)_{i}}
$$
and that of the other $U(1)$'s or nonabelian gauge fields to the same
R-R field $B^{\mu \nu}_{k}$,
$$
Tr(\gamma^{-1}_{k} \lambda^{2}_{G}) (\partial^{[ \mu} B^{\nu \rho
]}_{k}) W^{CS}_{\mu \nu \rho}
$$
This is because that these closed string low-energy amplitude will be
proportional to\cite{9808139}
\begin{equation} \label{eq: 77}
\sum_{k=1}^{3} C^{pq}_{k} Tr(\gamma_{k, p} \lambda^{(p)}_{i})
Tr[\,\gamma^{-1}_{k, q} (\lambda^{(q)}_{j} )^2\,]
\end{equation}

After the cancellation of $U(1)$ anomalies via above Green-Schwarz
mechanism, three of the four independent $U(1)$ charges become into
the global symmetries of the system, while the remaining one which
is equivalent to weak hypercharge
$$
Y= \frac{2}{3}Q_{3} + \frac{1}{2} Q_{2} + Q_{0}
$$
survives as an  (Abelian) gauge interaction.  Needless to say, this is
just what we expect. Because $Q_{3}$ can  be interpreted as the baryon
number, its conservation guarantees the stability of proton.

\section{Conclusions}
\renewcommand{\theequation}{5.\arabic{equation}}
\setcounter{equation}{0}

In this  paper, we have succeeded  getting the chiral fermion spectrum
of   standard  model   and  3-generation  replication  of quark-lepton
families from the $D=4$ Type IIB orbifold  $T^6/{\bf Z_3}$. A stack of
D-branes    (  D3-,   D7$_4$-   and    D7$_5$-  )  and   anti-D-branes
(${\widetilde{\textrm D7}}_3$)  are  necessary to  locate  at orbifold
singularities. The obtained model is non-supersymmetric but it is free
of  cubic $SU(3)$ gauge anomaly.  The  possible $U(1)$ gauge anomalies
can be   cancelled via a   generalized Green-Schwarz  mechanism. As  a
result,  the proton decay  is avoided by  the global ``baryon number''
conservation.

It is plausible that the  present paper provides a stringy realization
of the phenomenological  D-brane model of Ref.\cite{0004214}. There is
no unification for gauge coupling constants in this system. We suppose
that  the $U(1)_{j}$   charge is     measured  with respect   to   the
corresponding   coupling ${g_j}/{\sqrt{2j}}$,  with $g_j$ the  $SU(j)$
coupling constant. Then the expression for hypercharge leads to:
\begin{equation} \label{eq: 78}
\left. \begin{array}{lll}
\frac{1}{g^{2}_{Y}} & = & \frac{6}{g^{2}_{3}} (\frac{2}{3})^{2}
 + \frac{4}{g^{2}_{2}} (\frac{1}{2})^{2} + \frac{2}{g^{2}_{0}}\\
&    &  \\
&  = & \frac{8}{3 g^{2}_{3}} + \frac{1}{g^{2}_{2}} +
\frac{2}{g^{2}_{0}}
\end{array}
\right.
\end{equation}
This gives rise to the following expression for weak angle,
\begin{equation} \label{eq: 79}
\sin^{2}\theta_{W} \equiv \frac{g^{2}_{Y}}{g^{2}_{2} + g^{2}_{Y}}
\,=\, [\,2 + \frac{8}{3}(\frac{g_2}{g_3})^2 + 2 (\frac{g_2}{g_0})
\,]^{-1}
\end{equation}
In   order to compare  the above  $\sin^{2}\theta_{W}$ with low-energy
data, running from the string scale to  the weak scale should be taken
into account. Just as what  the authors of Ref.\cite{0004214} did,  we
can postulate  that the  coupling  constant $g_0$  is either  $g_3$ or
$g_2$.  In  this way, it seems  possible for our model  to accommodate
the low-energy phenomenology.

The non-supersymmetric D-brane orbifold under consideration is free of
the instability due  to the  brane-anti-brane  annihilation.  In  fact
there is no tachyon state  in its NS open  string spectrum. In view of
the analysis  of  Ref.\cite{9908072}, tachyons  only  appear when  the
model contains coincident or very  close branes and anti-branes of the
same  type, which  does  not happen in  our  orbifold.  Relying on the
absence  of tachyon states,   at string scale   the Higgs mass squared
would  be positive and the spontaneous  gauge symmetry breaking at the
electroweak  scale     would    be   triggered   by    the   radiative
corrections\cite{0012288}.  This  mechanism is  conjectured to be very
similar to the radiative electroweak  symmetry breaking of the minimal
supersymmetric standard  model(MSSM), which  will depend strongly upon
the  Yukawa  couplings and    the   structure  of  the   full   global
model.  Furthermore, the global    cancellation  of R-R  charges  (the
untwisted R-R tadpoles) may be  achieved by identifying the considered
${\bf Z_3}$ Type IIB  orbifold as a subspace of  ${\bf Z_6}$ Type  IIB
orientifold\cite{0010091}.    It is expected that   in the ${\bf Z_6}$
orientifold, which contains the ${\bf Z_3}$ orbifold fixed points, the
total R-R  charge of the branes  and anti-branes in  our present model
could  be   cancelled  by the  opposite  R-R   charge  carried  by the
orientifold planes. We will discuss these issues in the near future.

\subsection*{Acknowledgments}

H.X. Yang is grateful to Prof. David Bailin and Dr. Maria E.
Angulo for their help in many involved topics. Financial support
from Pao's scholarship for Chinese studying overseas is kindly
acknowledged.

\end{document}